\documentclass[preprint,aps]{revtex4}
\usepackage{graphicx}
\usepackage{dcolumn}
\begin{document}
\title{Purely-long-range bound states of He$(2s\,{}^3S)+$He$(2p\,{}^3P)$}

\author{V. Venturi, P. J. Leo, E. Tiesinga, and C. J. Williams}
\affiliation{Atomic Physics Division, National Institute of Standards and
Technology, Gaithersburg, Maryland 20899}

\author{I. B. Whittingham}
\affiliation{School of Mathematical and Physical Sciences, James Cook
University, Townsville 4811, Australia}

\date{\today}

\begin{abstract}
We predict the presence and positions of purely-long-range bound states of ${}^4$He$(2s\,{}^3S)+{}^4$He$(2p\,{}^3P)$
near the $2s\,{}^3S_1+2p\,{}^3P_{0,1}$ atomic limits. The results of the full multichannel and approximate models
are compared, and we assess the sensitivity of the bound states to atomic parameters characterizing the potentials.
Photoassociation to these purely-long-range molecular bound states
may improve the knowledge of the scattering length associated with the collisions of two ultracold spin-polarized
${}^4$He$(2s\,{}^3S)$ atoms, which is important for studies of Bose-Einstein condensates.
\end{abstract}

\maketitle

\section{Introduction}

Photoassociation spectroscopy to purely-long-range molecular bound states has proven a powerful tool for examining the ground state
wavefunction of alkali atoms, and for precisely determining $s$-wave scattering lengths for these
systems~\cite{Lettetal:1995,Stwalley&Wang:1999,Tiesingaetal:1996}. In recent years there has been a growing interest in rare-gas atoms,
and in particular helium systems. The production of Bose-Einstein condensates of spin-polarized metastable ${}^4$He atoms by two different
groups~\cite{Robertetal:2001,PereiraDosSantosetal:2001} allows one to study the properties of these condensates and provides an improved
environment for further investigation of cold collisions of metastable ${}^4$He~\cite{IOTA,ENS}. Additional valuable information about the
metastable ${}^4$He system may be obtainable from
photoassociation spectroscopy. Herschbach {\it et al.}~\cite{Herschbachetal:2000} observed bound
states that dissociate to the ${}^4$He$(2s\,{}^3S_1)+{}^4$He$(2p\,{}^3P_2)$ atomic limit by detecting ions produced by Penning ionization.
Broadening of the photoassociation peaks due to Penning ionization did not hinder the observation of the vibrational series.
There have been no published theoretical or experimental studies of the ${}^4$He$(2s\,{}^3S_1)+{}^4$He$(2p\,{}^3P_{0,1})$ purely-long-range
bound states.

In this paper we provide theoretical predictions for binding energies of levels of purely-long-range molecular potentials that
dissociate to the ${}^4$He$(2s\,{}^3S_1)+{}^4$He$(2p\,{}^3P_{0,1})$ atomic limits.  The purely-long-range bound states have inner turning points at internuclear separations larger
than $150\,a_0$ ($1a_0=5.29177\times10^{-11}$m) and primarily depend on the dipole-dipole $C_3$ coefficient and the fine structure
splittings of the ${}^3P$ atom.
These quantities are all determined from atomic properties.
The purely-long-range bound states are independent of the short-range form of the interaction potentials. Consequently, there may not
be a detectable ion signal from Penning ionization processes, which occur at small internuclear separations, $R<20\,a_0$.
We cannot make theoretical predictions for levels that have a short-range contribution,
because the multipole expansion of the potentials is not appropriate
for $R<30\,a_0$ and no short-range ${}^4$He$(2s\,{}^3S)+{}^4$He$(2p\,{}^3P)$ potentials exist.
For the remainder of the paper, we implicitly mean ${}^4$He when referring to helium.

\section{Method}

We calculate the molecular eigenfunctions of a multichannel Hamiltonian containing the relative kinetic energy operator, twelve
non-relativistic Born-Oppenheimer potentials which include retardation effects, the atomic spin-orbit interaction of the $^3P$ atom,
and a term describing nuclear rotation.
The model used is similar to that applied to the alkali systems~\cite{Tiesingaetal:1996}. The Movre-Pichler
model~\cite{Movre-Pichler:1977} adapted to ${}^3S+{}^3P$ systems yields the adiabatic potentials obtained from diagonalizing, at each
internuclear separation, the multichannel potentials that include the dipole-dipole and spin-orbit interactions.
In our calculations we have not included Penning ionization or spontaneous emission, which can both be represented as an optical or
imaginary potential. The radial extent of the optical potential for Penning ionization is localized well inside
the inner turning point of the purely-long-range bound states and therefore has a minor role to play for these states.
Spontaneous emission is present at all internuclear separations. However, near the $^3S$+$^3P$
dissociation limits the vibrational level spacing is large compared to the atomic spontaneous emission width of 1.63 MHz, and
spontaneous emission can therefore be treated perturbatively.

The multichannel Hamiltonian has been set up as described in Ref.~\cite{Singeretal:1983,Gao:1996}. The collision between two helium atoms
conserves the total molecular angular momentum $J$ and parity $p$. The total angular momentum is given by
$\mathbf{J}=\mathbf{l}+\mathbf{j}$, where $\mathbf{l}$ is the rotational angular momentum and $\mathbf{j}=\mathbf{L}+\mathbf{S}$.
Here $\mathbf{L}$ and $\mathbf{S}$ are the electronic orbital and spin angular momenta and $\Lambda$ and $\Sigma$ are their respective
projections onto the internuclear axis. The projection of $\mathbf{j}$ onto this axis is $\Omega=\Lambda+\Sigma$.
In addition, $\mathbf{j}=\mathbf{j_a}+\mathbf{j_b}$, where $\mathbf{j_a}$ and $\mathbf{j_b}$ are the total angular momenta of the
${}^3S$ and ${}^3P$ atoms, respectively.
Since ${}^4$He has zero nuclear spin, $p=+$ and $p=-$ parity states correspond to gerade ($g$) and ungerade ($u$) states,
respectively. The $\sigma=g/u$ labels refer to the inversion symmetry of the electron wavefunctions about the center of charge.
Negative-parity bound states have contributions from even partial waves $l$ only. Consequently, photoassociation of two ultracold metastable
helium atoms, where $s$-wave collisions dominate, can only excite ungerade ${}^3S+{}^3P$ bound states.

The twelve Born-Oppenheimer potentials are uniquely labelled by ${}^{2S+1}\Sigma_\sigma^+$ for $\Lambda=0$ or
${}^{2S+1}\Pi_\sigma$ for $|\Lambda|=1$, where the $+$ superscript denotes the symmetry of the spatial electronic wavefunction with respect
to reflection through a plane containing the internuclear axis. For $R>30 \, a_0$, the dispersion potentials are given by
$f_{3\Lambda}(R/\lambdabar) C_{3\Lambda}/R^3+ C_{6\Lambda}/R^6$, where $f_{3\Lambda}$ is an $R$- and $\Lambda$-dependent retardation
correction~\cite{Meath:1968}. The quantity $\lambdabar=3258.17 \, a_0$, where $\lambdabar=\lambda/(2\pi)$ and $\lambda$
is the  $2s\,{}^3S-2p\,{}^3P$ transition wavelength. The $C_{3\Sigma}$ coefficient is $\pm 2 C_3$ and $C_{3\Pi}$ is $\pm C_3$,
where $C_3$ is $6.41022 \, E_h a_0^3$~\cite{Drake_pc}, $E_h$ is a Hartree, and $C_3$ is proportional to the square of the
$2s\,{}^3S-2p\,{}^3P$  transition dipole matrix element.
Following the Physical Reference Database~\cite{NISTREF}, we assume a 1\% uncertainty or
``extent of the possible error'' for this coefficient. The $C_{3\Lambda}/R^3$ contributions to the potentials are attractive for the
${}^1 \Sigma_u^+$, ${}^3 \Sigma_g^+$, ${}^5 \Sigma_u^+$, ${}^1 \Pi_g$, ${}^3 \Pi_u$, and ${}^5 \Pi_g$ states, and repulsive for
the ${}^1 \Sigma_g^+$, ${}^3 \Sigma_u^+$, ${}^5 \Sigma_g^+$, ${}^1 \Pi_u$, ${}^3 \Pi_g$, and ${}^5 \Pi_u$ states.
The van der Waals coefficients are $C_{6\Sigma}=2620.76$ $E_h a_0^6$ and $C_{6\Pi}=1846.60$ $E_h a_0^6$~\cite{Marinescu_private_comm}. These
can be calculated from atomic dipole moments and transition frequencies~\cite{Kharchenkoetal:1997}.
In turns out that the $C_3/R^3$ term is the dominant contribution to the dispersion potential relevant for the purely-long-range states.

The $j_b$ quantum number for the $2p \, {}^3P$ atom, which describes the spin-orbit interaction, has the values 0,1, or 2.
The $2p \, {}^3P_0$ and $2p \, {}^3P_1$ levels lie $31.9081$ GHz and $2.2912$ GHz above the $2p \, {}^3P_2$ level,
respectively~\cite{Castillegaetal:2000,Georgeetal:2001}.
Note that these energy splittings do not satisfy the Land\'e intervals. The rotational Hamiltonian is given by $\hbar^2\mathbf{l}^2/2\mu R^2$,
where $\mu=m/2$ is the reduced mass of the system. The atomic mass is $m=4.0026032$ u, where 1 u $= 1.66053873\times10^{-27}$ kg.

We investigate the sensitivity of the bound states to the short-range Born-Oppenheimer potentials. No data on the short-range shape currently
exists, so we use two different forms for the short-range potentials for $R<30\,a_0$. The first of these short-range forms assumes a hard wall
for all twelve potentials at $R=30\,a_0$. For smaller internuclear separations the potentials are no longer described by the
dispersion potential. For the second short-range form, the attractive Born-Oppenheimer potentials are given a more realistic
Lennard-Jones-like form, with inner turning points between $4 \, a_0$ and $5 \, a_0$.

The molecule is best described by the Hund's case (a) coupling scheme at small $R$, where the splitting between the Born-Oppenheimer potentials
is large compared to the spin-orbit interaction. The molecular states are well described by the quantum numbers
$J$ and ${}^{2S+1}\Lambda_\sigma^\pm$. As the internuclear separation increases and the
energy separation between the Born-Oppenheimer curves becomes comparable to the atomic spin-orbit interaction, the
Hund's case (a) coupling scheme becomes less
appropriate, and the $R$-dependent Hund's case (c) coupling scheme is a more suitable choice for describing the long-range states
and potentials.
The Hund's case (c) molecular states are described by the quantum numbers $J$, $\Omega_\sigma^\pm$. For $\Omega=0$ the
$\pm$ superscript label denotes the symmetry of the electronic wavefunction with respect to reflection through a plane
containing the internuclear axis. We note that for the $\Omega=0$ states, only $0^-_u$ and $0^+_g$ exist for even $J$,
while the $0^+_u$ and $0^-_g$ states exist for odd $J$. The Hund's case (c) states are a linear
combination of the Hund's case (a) states. Table~\ref{tab:states} gives the Hund's case (a) states that contribute to a
given Hund's case (c) state.

The purely-long-range states discussed in this paper are in the radial regime described by the Hund's case (c) coupling scheme.
We will discuss various approximations to the full multichannel Hamiltonian used in calculating these states.
The adiabatic Movre-Pichler model described previously labels each purely-long-range state by $\Omega_\sigma^\pm$.
An improvement on this model is the multichannel Hamiltonian that includes the Born-Oppenheimer potentials and the spin-orbit interaction.
In addition, we discuss two ways of including the nuclear rotation.
These approximate Hamiltonians are useful for assigning vibrational quantum numbers to the bound states calculated using
the full multichannel model, and the adiabatic Movre-Pichler model is useful for confirming the assignment of Hund's case (c) labels to the
bound states.

\begin{figure}[hb]
  \begin{center}
    \includegraphics[width=0.5\linewidth]{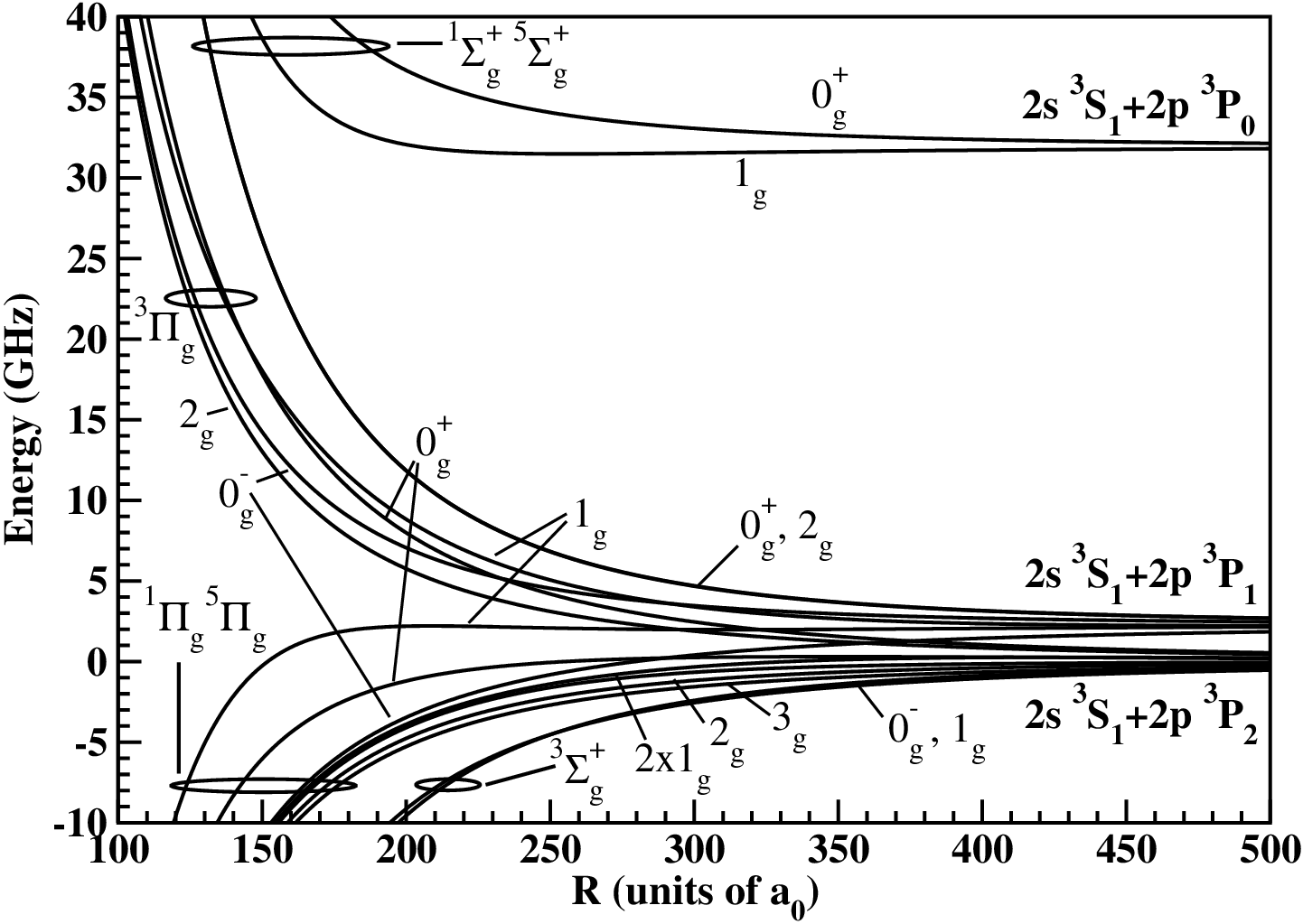}
  \end{center}
  \caption{Adiabatic extended Movre-Pichler potentials of gerade symmetry as a function of internuclear
separation. The zero of energy is the He$(2s\,{}^3S_1)+$He$(2p\,{}^3P_2)$ atomic limit.
The potentials are labelled by $\Omega_g^\pm$. At short-range we have indicated with ellipses the ${}^{2S+1}\Lambda_g^+$ correlations.
}
  \label{fig:potg}
\end{figure}

\begin{figure}[hb]
  \begin{center}
    \includegraphics[width=0.5\linewidth]{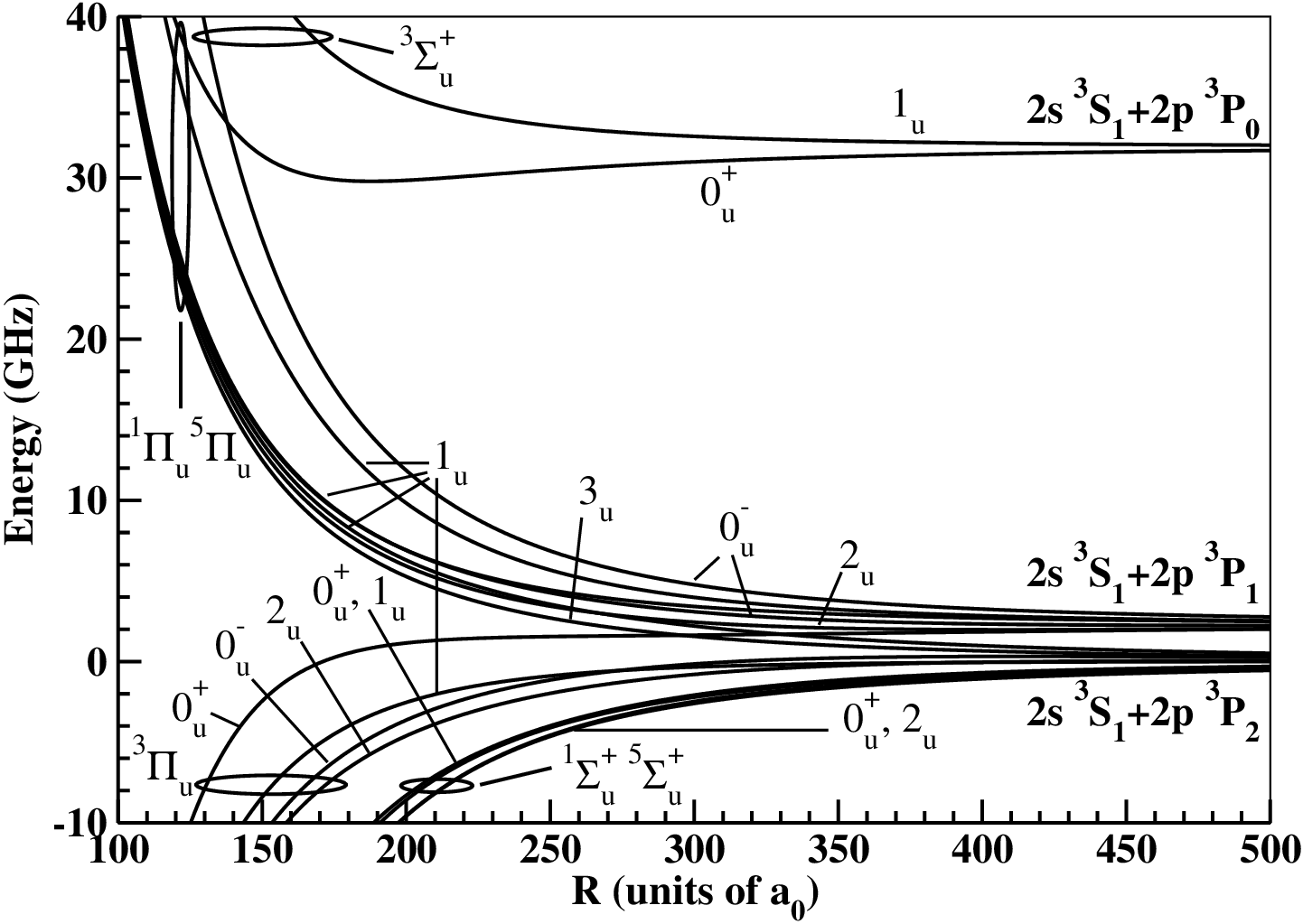}
  \end{center}
  \caption{Adiabatic extended Movre-Pichler potentials of ungerade symmetry as a function of internuclear
separation. The zero of energy is the He$(2s\,{}^3S_1)+$He$(2p\,{}^3P_2)$ atomic limit.
The potentials are labelled by $\Omega_u^\pm$. At short-range we have indicated with ellipses the ${}^{2S+1}\Lambda_u^+$ correlations.
}
  \label{fig:potu}
\end{figure}

The adiabatic Movre-Pichler potentials are shown in Figs.~\ref{fig:potg} and~\ref{fig:potu}. They are obtained using an extension of the
model defined in Ref.~\cite{Movre-Pichler:1977}. This extended Movre-Pichler model is adapted to the ${}^3S+{}^3P$ system and,
in addition to the usual dipole-dipole and the spin-orbit interactions, includes the $C_6/R^6$ dispersion term and retardation.
As mentioned previously, the fine structure splittings for the He$(2p \, {}^3P)$ atom do not satisfy the Land\'e intervals, which results
in a coupling between molecular singlet and quintet states. For large $R$, the potentials in
Figs.~\ref{fig:potg} and~\ref{fig:potu} dissociate to the three ${}^3P$ fine structure limits.
For smaller $R$ the potentials correlate to attractive and repulsive Born-Oppenheimer
potentials, as indicated in the figures. We find a number of purely-long-range potential wells in the adiabatized potentials dissociating to
the $2s\,{}^3S_1+2p\,{}^3P_0$ and $2s\,{}^3S_1+2p\,{}^3P_1$ atomic limits
that can support bound states. In Figs.~\ref{fig:potg} and \ref{fig:potu} the $1_g$ and $0_u^+$ states dissociating to
the $2s\,{}^3S_1+2p\,{}^3P_0$ limit provide an example of such wells, while the wells in the states dissociating to
the $2s\,{}^3S_1+2p\,{}^3P_1$ limit are too shallow to be observed on the scale of the present figure. There are no purely-long-range
states dissociating to the $2s\,{}^3S_1+2p\,{}^3P_2$ limit.

The eigenstates of the Hamiltonian are determined by multichannel bound state calculations in a manner similar to that described in
Ref.~\cite{Tiesingaetal:1996}. Long-range bound states are not ``true'' bound states of the system when
their energy lies above the $2s\,{}^3S_1+2p\,{}^3P_2$ dissociation limit. They are bound states embedded in a continuum and consequently can
predissociate. In our method a hard wall is placed at a large internuclear separation, $R>5000\,a_0$. Consequently,
the calculated states are either bound states or ``box states'', which characterize the continuum.
These two types of states are distinguished by visual inspection of a multichannel eigenvector
$|\Psi(R)\rangle=\sum_i \psi_i(R)|i\rangle$ or the square root of the density $(\sum_i |\psi_i(R)|^2)^{1/2}$, where
the sum is over a finite number of molecular spin states $|i\rangle$. The function $\psi_i$ is the radial component for spin state
$|i\rangle$, which is characterized by the total spin, parity, and other angular momenta.
The two types of states can also be distinguished by their sensitivity to the location of the hard wall boundary condition.
The predissociation of the bound states will be further discussed in the next section.

\section{Results}

Four sets of purely-long-range ro-vibrational states are found in the multichannel analysis and the binding energies for these levels are
given in Table~\ref{tab:bound_states}. We report here only levels with binding energies greater than 1 MHz even though states bound by less
than twice the atomic line width, $3.25$ MHz, may not be clearly observable. The levels are characterized using Hund's case (c) labels.
The $1_g$ and $0_u^+$ potentials dissociating to the
$2s\, {}^3S_1+2p\,{}^3P_0$ limit and the $0_u^-$ and $2_u$ potentials dissociating to the $2s\, {}^3S_1+2p\,{}^3P_1$ limit have
purely-long-range states. Each of these potentials has no more than 6 vibrational levels. These levels are labelled by $v$ and exhibit a
rotational progression when $J\ge\Omega$. In addition, $0_u^+$ $(0_u^-)$ states only exist for $J$ odd (even).
The uncertainty in the binding energy due to changing the short-range Born-Oppenheimer potentials is negligible.
The last significant digit of the binding energy indicates the predissociation width, an estimate of which is obtained by comparing the
energies for varying locations of the hard wall at large $R$. We find the predissociation width is on the order of, or smaller than, the
natural line width. Gerade/ungerade symmetry states can only be formed by photoassociation via odd/even partial wave collisions. At the
internuclear separations of relevance to the states described here, retardation corrections that introduce dipole-forbidden coupling between
gerade and ungerade states~\cite{Machholmetal:2001} are small.

As shown in Table~\ref{tab:bound_states} for the lowest allowed rotational level of each state,
the 1\% extent of the error on $C_3$ produces a change of less than 1 MHz for the $1_g$ bound states dissociating to the
$2s\, {}^3S_1+2p\,{}^3P_0$ limit and the $0_u^-$ and $2_u$ bound states dissociating to the $2s\, {}^3S_1+2p\,{}^3P_1$ limit. For the
$0_u^+$ bound states dissociating to the $2s\, {}^3S_1+2p\,{}^3P_0$ limit, changes of up to 4 MHz are found for the $v=0$ to $v=3$ levels.
The higher vibrational levels are shifted by less than 1 MHz. These changes are obtained using a fixed location of the hard wall at large $R$.
It is necessary to include retardation as it leads to a small correction
in the binding energy. For the deepest purely-long-range potential, the $0_u^+$ dissociating to the
$2s\, {}^3S_1+2p\,{}^3P_0$ limit, the retardation correction lowers the bound state energy of the $J=1$, $v=0$ ro-vibrational level by
approximately 7 MHz. The complete $C_6/R^6$ contribution to the bottom of this  $0_u^+$ well at $R\approx 190 \, a_0$  is
approximately 0.4 MHz. Therefore the shift in the binding energy due to any uncertainty in $C_6$ is small compared with the natural line width.

\begin{figure}[hb]
  \begin{center}
    \includegraphics[width=0.5\linewidth]{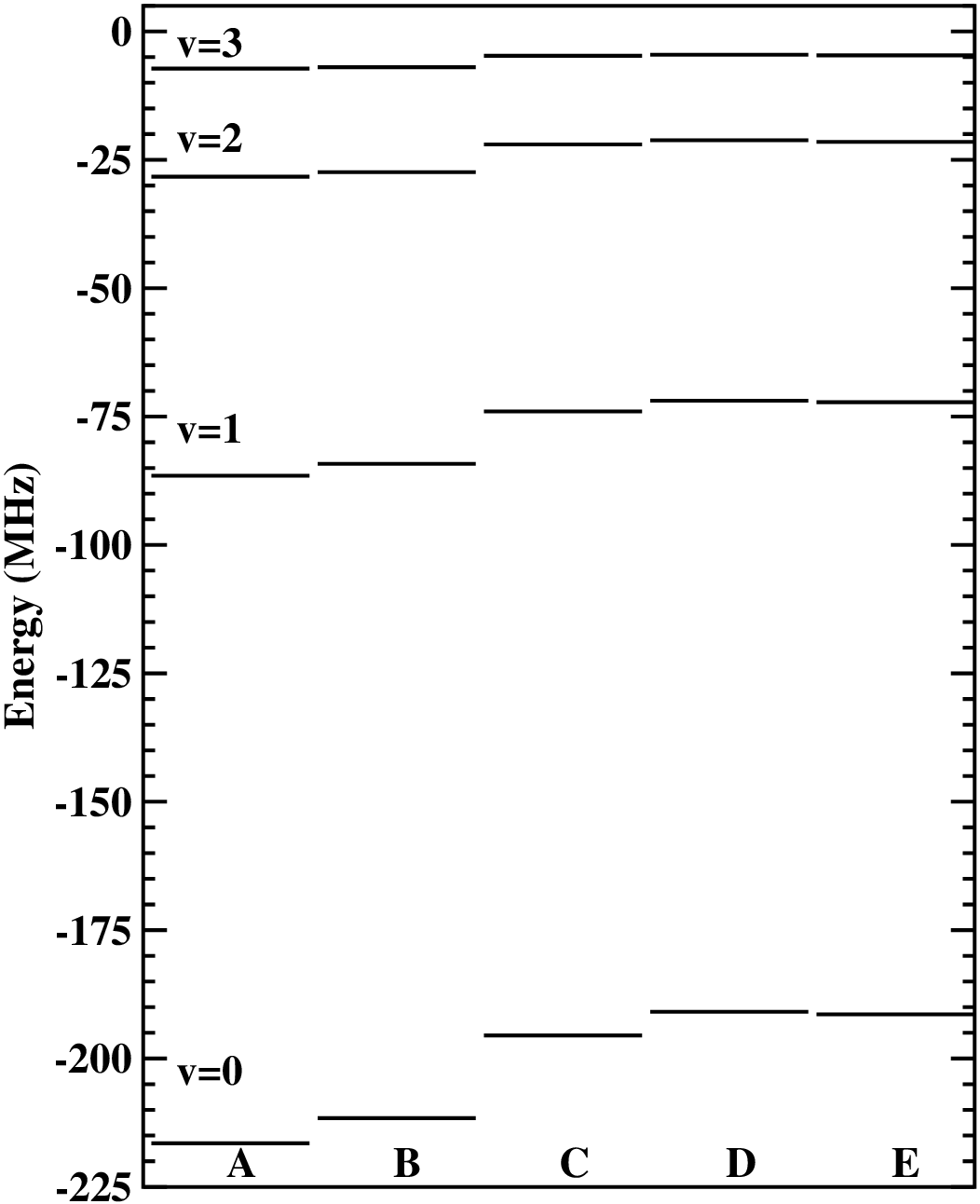}
  \end{center}
  \caption{The binding energy of the $v=0$ to $v=3$ vibrational levels of the purely-long-range $2_u$ potential dissociating to the
  He$(2s\, {}^3S_1)+$He$(2p\, {}^3 P_1)$ limit. Various approximations to the complete multichannel model
  are shown. Working from a basic model and increasing in complexity to the complete multichannel calculation:
  (A) the adiabatic extended Movre-Pichler approximation, (B) a multichannel calculation including the Born-Oppenheimer potentials and the
  atomic spin-orbit interaction, (C) an adiabatic calculation
  of the full $J=2$ Hamiltonian where the Coriolis coupling between different $\Omega_u^\pm$ states has been set to zero,
  (D) a multichannel calculation of the full $J=2$ Hamiltonian where the Coriolis coupling between different $\Omega_u^\pm$ states has been set
  to zero, and (E) the complete $J=2$ multichannel calculation.
  }
  \label{fig:coupling}
\end{figure}

We look at three examples to clarify the ro-vibrational and Hund's case (c) assignments of the levels reported in Table~\ref{tab:bound_states}.
In Fig.~\ref{fig:coupling} we show the binding energies for $J=2$, $v=0$ to $v=3$ ro-vibrational levels of the $2_u$ state dissociating to the
He$(2\, {}^3S_1)+$He$(2\, {}^3 P_1)$ limit for four approximations and the exact multichannel calculation. The four approximations are:
(A) the adiabatic extended Movre-Pichler approximation, (B) a non-rotating multichannel calculation including the Born-Oppenheimer potentials
and the atomic
spin-orbit interaction, (C) an adiabatic calculation of the full $J=2$ Hamiltonian where the Coriolis coupling between different $\Omega_u^\pm$
states has been set to zero, and (D) a multichannel calculation of the full $J=2$ Hamiltonian where the Coriolis coupling between different
$\Omega_u^\pm$ states has been set to zero. The binding energies labelled by (E) are obtained using a multichannel calculation of the
full $J=2$ Hamiltonian. All four approximations predict the correct binding energies to within 10\%. Experiments are
sufficiently accurate that the full multichannel model must be used. For the $v=0$ level, the 4.9 MHz difference between approximations (A) and
(B) is due to the positive diagonal non-adiabatic coupling, which depends on the Born-Oppenheimer and spin-orbit
interactions~\cite{Jonesetal:1996}. The 21 MHz discrepancy between the two adiabatic approximations (A) and (C)
for the $v=0$ level can qualitatively be understood by the perturbative inclusion of the $\hbar^2\mathbf{l}^2/(2\mu R^2)$ rotational
Hamiltonian. For this Hund's case (c) $2_u$ state the $\mathbf{l}^2$ of the rotational Hamiltonian simplifies to
$J(J+1)-2\Omega^2+j(j+1)=J(J+1)-2$, since $j=\Omega=2$ near the He$(2\, {}^3S_1)+$He$(2\, {}^3 P_1)$ dissociation limit. Clearly, for the lowest
rotational level $J=2$, a non-zero rotational correction occurs. A non-pertubative calculation is needed for a quantitative understanding of
the discrepancy between the two adiabatic approximations (A) and (C) as rotational distortions are important. Similar to the difference
between approximations (A) and (B), the difference between approximations (C) and (D) is also due to a positive diagonal non-adiabatic
coupling, which now depends on the rotational Hamiltonian as well as the Born-Oppenheimer and spin-orbit interactions. In this case,
the comparison of approximation (D) with (E) shows that the coupling between $\Omega$ states shifts the $J=2$, $2_u$ levels by no more than
0.5 MHz. Similar comparisons can be obtained for all of the purely-long-range states.

\begin{figure}[hb]
  \begin{center}
    \includegraphics[width=0.5\linewidth]{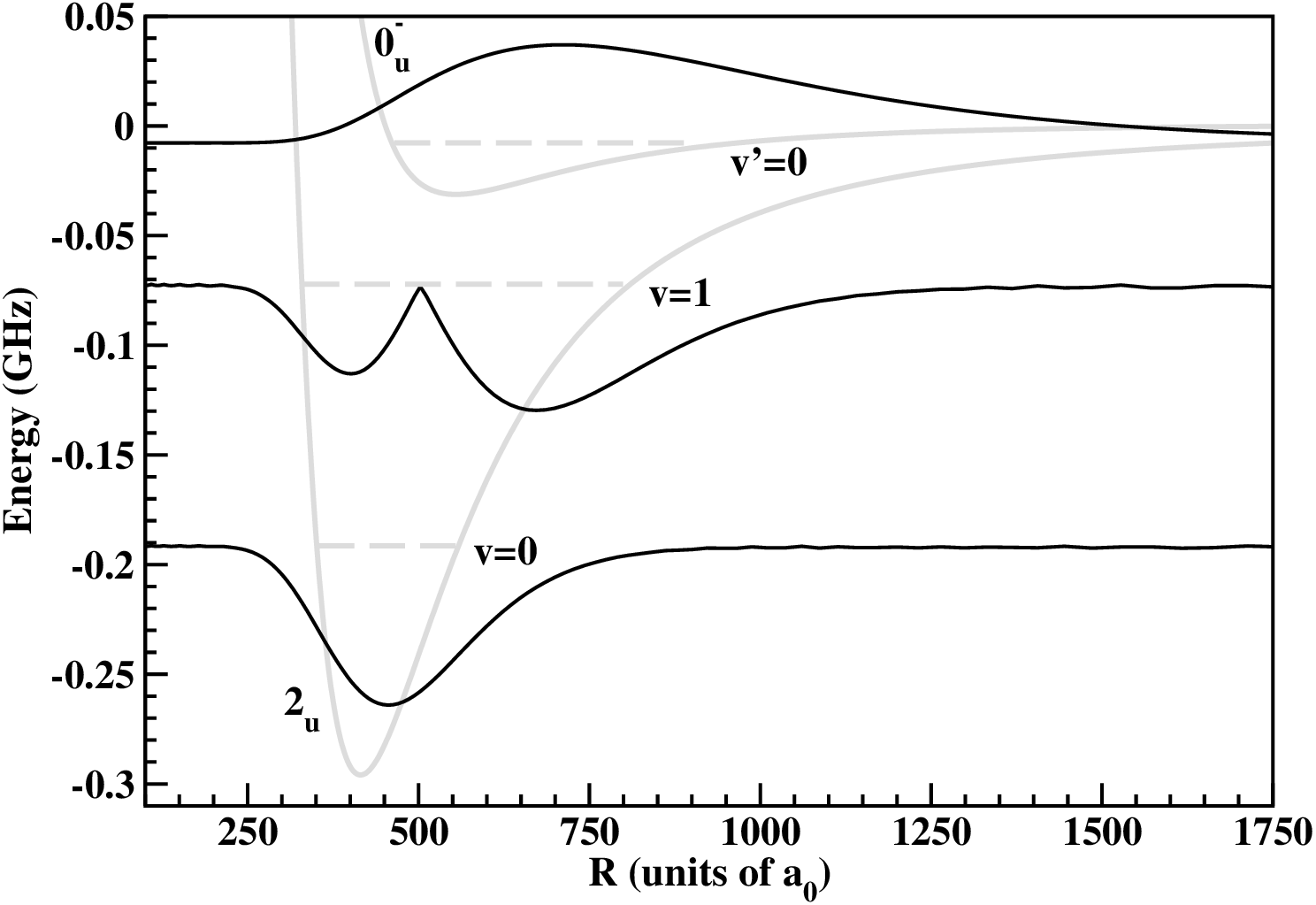}
  \end{center}
  \caption{The square root of the density for the $J=2$, $v=0$ and $v=1$ ro-vibrational levels of the $2_u$ state and the $J=2$,
  $v'=0$ ro-vibrational level of the $0_u^-$ state. Note that for the lower two levels the square root of the density curves have been
  inverted to maintain the distinguishability of the curves. The solid grey curves represent
  the adiabatic $J=2$, $2_u$ and $0_u^-$ potentials dissociating to the He$(2s\, {}^3S_1)+$He$(2p\, {}^3 P_1)$ limit.
  The binding energies of the levels are illustrated by dashed horizontal lines.
 }
  \label{fig:levels_nsr}
\end{figure}

In Fig.~\ref{fig:levels_nsr} we show the square root of the density for exact $J=2$ multichannel wavefunctions assigned with the
$2_u$ and $0_u^{-}$ symmetries. Also shown are the $2_u$ and $0_u^{-}$ adiabatic potentials dissociating to the $2s\, {}^3S_1+2p\,{}^3P_1$
limit, as obtained using the approximation (C) described previously. These bound levels are purely-long-range because the square roots
of the densities show no amplitude for $R<250 \, a_0$. In addition, the square roots of the densities calculated using the full multichannel
and adiabatic models do not deviate appreciably. A clear vibrational progression from the $v=0$ to the $v=1$ level of the
$2_u$ state is visible.

\begin{figure}[hb]
  \begin{center}
    \includegraphics[width=0.5\linewidth]{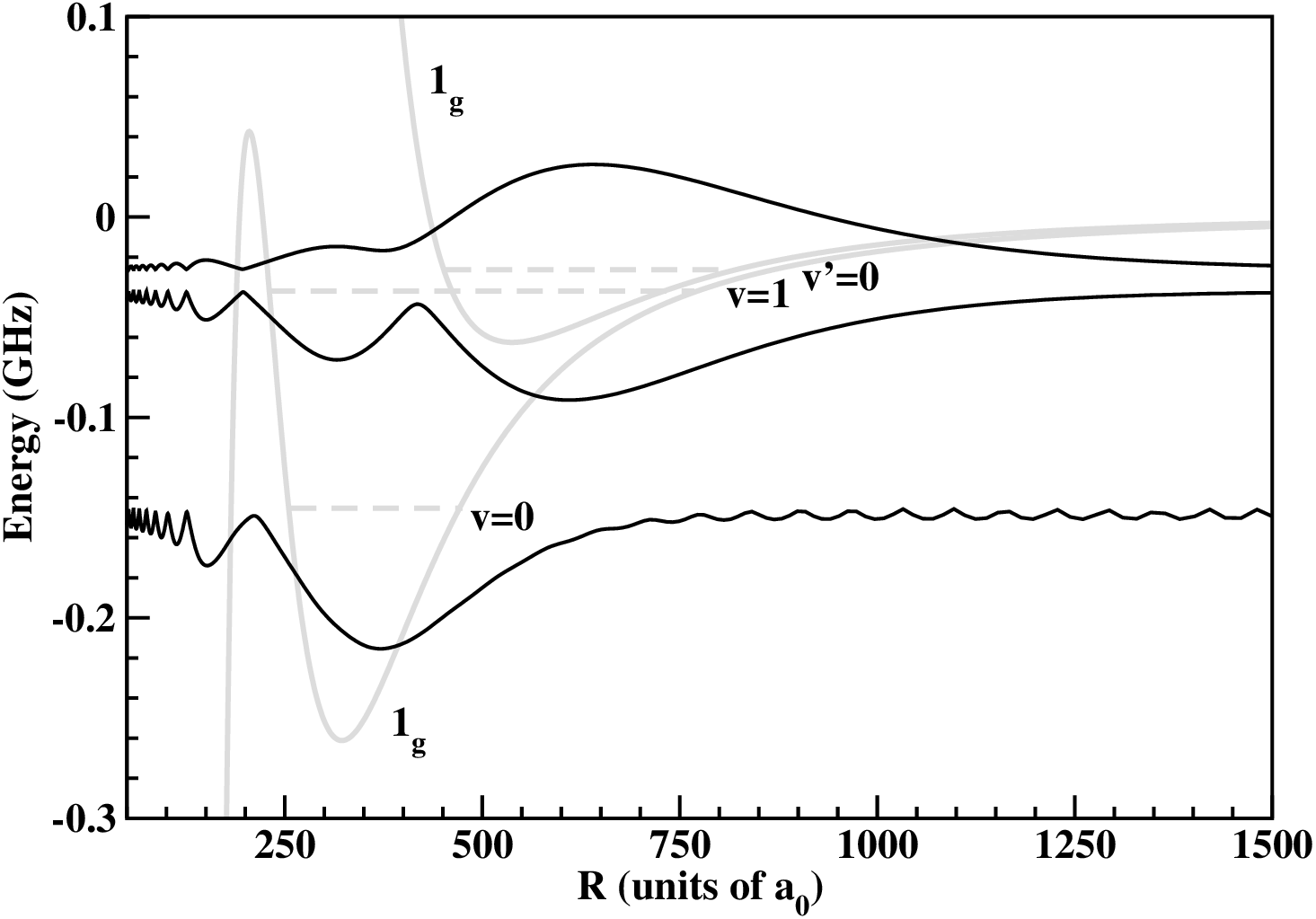}
  \end{center}
  \caption{The square root of the density for the $J=1$, $v=0$ and $v=1$ ro-vibrational levels of the deep $1_g$ state and the $J=1$, $v'=0$
  ro-vibrational level of the shallow $1_g$ state. Note that for the lower two levels the square root of the density curves have been
  inverted to maintain the distinguishability of the curves. The solid grey curves represent the adiabatic $J=1$, $1_g$ potentials dissociating to the
  He$(2s\, {}^3S_1)+$He$(2p\, {}^3 P_1)$ limit. The density shows significant amplitude for internuclear separations less than $200 \, a_0$.
  The binding energies of the levels are illustrated by dashed horizontal lines.}
  \label{fig:levels_sr}
\end{figure}

The square root of the density calculated using the full Hamiltonian for three $J=1$, $1_g$ bound levels and two $J=1$, $1_g$ adiabatic
potentials dissociating to the $2s\, {}^3S_1+2p\,{}^3P_1$ limit are shown in Fig.~\ref{fig:levels_sr}. The two potentials are calculated using
approximation (C). The ``deep'' $1_g$ potential has a double-well structure with a minimum at $\approx 315\,a_0$.
The other ``shallow''  potential has a minimum at $\approx 530\,a_0$.
These levels have not been tabulated in Table~\ref{tab:bound_states} because they are sensitive to the
short-range potentials. We present them here as they are seemingly purely-long-range except for the penetration of the wavefunction
through the barrier of the double-well potentials. The peak of the barrier lies slightly above the $2s\, {}^3S_1+2p\,{}^3P_1$ dissociation
limit. The ``shallow'' potential is not a purely-long-range potential because of significant non-adiabatic mixing with the ``deep'' $1_g$
potential. This short-range dependence leads to an uncertainty in the level locations and possible broadening by Penning ionization
processes. In addition, the square root of the density for the $v=0$ level has an oscillatory behavior for $R>700 \, a_0$. The periodicity of
these oscillations corresponds to the periodicity of the square root of the density for box states in close proximity to the $v=0$ level.
For these $1_g$ states the amplitude of the oscillations is primarily determined by the rotational coupling. These oscillations indicate
significant predissociation.

\section{Conclusion}

We have predicted the binding energies of all four sets of purely-long-range ro-vibrational levels of ${}^4$He$(2s\,{}^3S)+{}^4$He$(2p\,{}^3P)$
potentials dissociating to the $2s\,{}^3S_1+2p\,{}^3P_{0,1}$ fine structure limits.
A comparison of the results calculated using the full multichannel and approximate models demonstrates that the full multichannel Hamiltonian
is required for accurate comparison with experiment. Retardation and the van der Waals $C_6$ interaction are found to give rise to small
corrections in the binding energies. We show that a measurement of the binding energy to at least 1 MHz leads to a 1\% determination of the
dipole-dipole $C_3$ coefficient.

The current measurements of the ${}^5\Sigma_g^+$ scattering length of colliding metastable ${}^4$He atoms are
$300\pm 150 \, a_0$~\cite{PereiraDosSantosetal:2001} and $380\pm 190 \, a_0$~\cite{Robertetal:2001}.
This implies that a node in the scattering wavefunction can be found between $150 \, a_0$ and $600 \, a_0$ ~\cite{Tiesingaetal:1996}.
The inner and outer turning points of our purely-long-range bound states
approximately fall within this range of internuclear separations. Consequently the intensity patterns as a function of vibrational level in
photoassociation spectroscopy may provide an improved measurement of the scattering length.

The experimental and theoretical binding energies for the $0_u^+$ state dissociating to the $2s\,{}^3S_1+2p\,{}^3P_{1}$ limit
reported in Ref.~\cite{Leonardetal:2003} came to our attention upon the completion of this paper. Our theoretical predictions
are within the experimental uncertainty for the $v=1-4$ levels and for the $v=0$ level our theoretical prediction of $-1418.1$ MHz is
2 MHz greater than the upper bound on the experimental binding energy $-1430\pm10$ MHz.

\bibliographystyle{prsty}

\begin{table}
\caption{The combinations of Hund's case (a) states which contribute to a
given case (c) state, for a particular $\sigma=g/u$ symmetry.}
\label{tab:states}
\begin{tabular}{cc}
case (c) state    &  case (a) states \\
\hline
$ 3_{\sigma} $    &  ${}^5\Pi_{\sigma}$ \\
$ 2_{\sigma} $    &  ${}^5\Pi_{\sigma}$, ${}^3\Pi_{\sigma}$,
		     ${}^5\Sigma_{\sigma}^+$ \\
$ 1_{\sigma} $    &  $2 \times ({}^5\Pi_{\sigma})$, ${}^3\Pi_{\sigma}$,
                     ${}^1\Pi_{\sigma}$, ${}^5\Sigma_{\sigma}^+$,
                     ${}^3\Sigma_{\sigma}^+$  \\
$ 0^+_{\sigma} $  &  ${}^5\Pi_{\sigma}$, ${}^3\Pi_{\sigma}$,
                     ${}^5\Sigma_{\sigma}^+$, ${}^1\Sigma_{\sigma}^+$  \\
$ 0^-_{\sigma} $  &  ${}^5\Pi_{\sigma}$, ${}^3\Pi_{\sigma}$,
                     ${}^3\Sigma_{\sigma}^+$  \\
\end{tabular}
\end{table}

\pagebreak

\begin{table}
\caption{Purely-long-range bound state energies in units of MHz, relative to the dissociation limit of the Hund's case (c) state.
$v$ and $J$ are the vibrational and rotational quantum numbers. The bound states for $J\le3$ are calculated, and the blank columns
indicate that the states do not exist for particular $J$.
The last column, $\Delta E_v$, indicates the change in binding energy of the
lowest rotational level for a +1\% change in the $C_3$ coefficient. All binding energies are calculated using the full multichannel
model.  }
\label{tab:bound_states}
\begin{tabular}{c|c| D{.}{.}{-1}  D{.}{.}{-1}  D{.}{.}{-1}  D{.}{.}{-1} D{.}{.}{-1}}
Symmetry$\rightarrow$dissociation limit & $ v $ & \multicolumn{1}{c}{$J=0$} & \multicolumn{1}{c}{$J=1$} & \multicolumn{1}{c}{$J=2$} &
\multicolumn{1}{c}{$J=3$} & \multicolumn{1}{c}{$\Delta E_v$}  \\
\hline \hline
$1_g \rightarrow 2s\,{}^3S_1+2p \, {}^3P_0$ & 0 & & -207.66 & -170.92 & -115.73 & 0.62 \\
      &1 & & -43.80 & -28.52 & -7.97 & 0.47 \\
      &2 & & -4.12 &        &        & 0.13 \\    \hline
$0_u^+ \rightarrow 2s\,{}^3S_1+2p \, {}^3P_0$  & 0 & & -1418.1 & & -1212.7 & 2.4 \\
& 1 & & -649   & & -513   & 3   \\
& 2      & & -253.12 & & -174.5 & 2.5  \\
& 3      & & -79.65 & & -41.6  & 1.35 \\
& 4      & & -18.30 & & -4.6   & 0.53 \\
& 5      & & -2.59  & &        & 0.13 \\     \hline
$0_u^{-} \rightarrow 2s\,{}^3S_1+2p \, {}^3P_1$  & 0 & -18.27 &  & -7.77 & & 0.08 \\
\hline
$2_u \rightarrow 2s\,{}^3S_1+2p \, {}^3P_1$ & 0 & & & -191 & -167 & 0.5  \\
      & 1 & & & -72 & -57 & 0.6 \\
      & 2 & & & -21.5 & -14.4 & 0.2 \\
      & 3 & & & -4.7  & -2.2  & 0.1 \\ \hline
\end{tabular}
\end{table}

\pagebreak



\end{document}